\documentclass[letter]{aa}  \usepackage[pdftex,pdfpagemode={UseOutlines},bookmarks,bookmarksopen,colorlinks,linkcolor={blue},citecolor={blue},urlcolor={red}]{hyperref}
\usepackage{natbib}
\usepackage{subfig}
\usepackage{graphicx}
\usepackage{txfonts}
\usepackage{array}
\usepackage{gensymb}
\usepackage{caption}

\newcolumntype{\AA}{AA}

\begin{document} 

   \title{The EBLM project\thanks{The data is publicly available at the CDS Strasbourg and on demand from the first author.}}
   \subtitle{III. A Saturn-size low-mass star at the hydrogen-burning limit}

\author{Alexander von Boetticher\inst{1,2}
\and Amaury H.M.J. Triaud\inst{2}
\and Didier Queloz\inst{1,3}
\and Sam Gill\inst{4}
\and Monika Lendl\inst{5,6}
\and Laetitia Delrez\inst{1, 9}
\and David R. Anderson\inst{4}
\and Andrew Collier Cameron\inst{7}
\and Francesca Faedi\inst{8}
\and Micha\"el Gillon\inst{9}
\and Yilen G\'omez Maqueo Chew\inst{10}
\and Leslie Hebb\inst{11}
\and Coel Hellier\inst{4}
\and Emmanu\"el Jehin\inst{9}
\and Pierre F.L. Maxted\inst{4}
\and David V. Martin\inst{3}
\and Francesco Pepe \inst{3}
\and Don Pollacco\inst{8}
\and Damien S\'egransan \inst{3}
\and Barry Smalley\inst{4}
\and St\'ephane Udry \inst{3}
\and Richard West\inst{8}
}

\offprints{av478@cam.ac.uk}

\institute{Cavendish Laboratory, J J Thomson Avenue, Cambridge, CB3 0HE, UK
\and Institute of Astronomy, Madingley Road, Cambridge CB3 0HA, UK
\and Observatoire Astronomique de l'Universit\'e de Gen\`eve, Chemin des Maillettes 51, CH-1290 Sauverny, Switzerland
\and Astrophysics Group, Keele University, Staffordshire, ST55BG, UK
\and Space Research Institute, Austrian Academy of Sciences, Schmiedlstr. 6, 8042, Graz, Austria
\and Max Planck Institute for Astronomy, K\"onigstuhl 17, 69117 Heidelberg, Germany 
\and SUPA, School of Physics \& Astronomy, University of St Andrews, North Haugh, KY16 9SS, St Andrews, Fife, Scotland, UK
\and Department of Physics, University of Warwick, Coventry CV4 7AL, UK
\and Universit\'e de Li\`ege, All\'ee du 6 ao\^ut 17, Sart Tilman, 4000, Li\`ege 1, Belgium
\and Instituto de Astronom\'ia, Universidad Nacional Aut\'onoma de M\'exico, Ciudad Universitaria, Ciudad de M\'exico, 04510, M\'exico
\and Hobart and William Smith Colleges, Department of Physics, Geneva, NY 14456, USA
}

   \date{12. June, 2017}

\abstract{
	We report the discovery of an eclipsing binary system with mass-ratio q $\sim$ 0.07. After identifying a periodic photometric signal received by WASP, we obtained CORALIE spectroscopic radial velocities and follow-up light curves with the {\it Euler} and TRAPPIST telescopes. From a joint fit of these data we determine that EBLM~J0555-57 consists of a sun-like primary star that is eclipsed by a low-mass companion, on a weakly eccentric 7.8-day orbit. Using a mass estimate for the primary star derived from stellar models, we determine a companion mass of $85 \pm 4 M_{\rm Jup}$ ($0.081M_{\odot}$) and a radius of $0.84^{+0.14}_{-0.04} R_{\rm Jup}$ ($0.084 R_{\odot}$) that is comparable to that of Saturn. EBLM~J0555-57Ab has a surface gravity $\log g_\mathrm{2} = 5.50^{+0.03}_{-0.13}$ and is one of the densest non-stellar-remnant objects currently known. These measurements are consistent with models of low-mass stars.
}
  
   \keywords{binaries: eclipsing; spectroscopic -- Stars: low-mass -- Stars: EBLM~J0555-57Ab -- techniques: spectroscopic; photometry}

\titlerunning{short title}
\authorrunning{name(s) of author(s)}
\maketitle  

Eclipsing binary stars enable empirical measurements of the stellar mass-radius relation. The low-mass regime, down to the hydrogen-burning mass limit, is poorly constrained by measurements of mass and radius, but is of particular relevance to the study of exoplanets. Stars with masses below $0.25 M_\odot$ are the most common stellar objects \citep{Kroupa_2001,Chabrier_2003, Henry_2006} and prove to be excellent candidates for the detection of Earth-sized planets \citep{Zachory_2015,Gillon_2016,Gillon_2017,Luger_2017} and their atmospheric characterization \citep{deWit_2016}.
Determining the properties of exoplanets requires an accurate knowledge of their host star parameters, in particular the stellar mass. This motivates the study of low-mass eclipsing binaries (henceforth EBLMs) \citep{EBLM_1, EBLM_II}, to empirically measure the mass-radius relation.
In this context, we report our results on the eclipsing binary EBLM~J0555-57. The system was detected by the Wide Angle Search for Planets \citep[WASP; \href{http://wasp-planets.net}{wasp-planets.net};][]{Pollacco_2006}, and was identified as a non-planetary false-positive through follow-up measurements with the CORALIE spectrograph. We use radial velocities and two eclipse observations by the TRAPPIST and \textit{Euler} telescopes, to determine the mass and radius of EBLM~J0555-57Ab, to 85.2$^{+4.0}_{-3.0}$~M$_{\rm Jup}$ (0.081M$_{\odot}$) and 0.84$^{+0.14}_{-0.04}$~R$_{\rm Jup}$ (0.084R$_{\odot}$). This places EBLM~J0555-57Ab at the minimum of the stellar mass-radius relation.  

\section{Observations}

The source 1SWASPJ055532.69-571726.0 (EBLM~J0555-57, J0555-57 for brevity) was observed by WASP-South between 2008-09-29 and 2012-03-22. The \textit{Hunter} algorithm \citep{Collier_Cameron_2007} detected 17 transit-like signals from 34~091 observations over four seasons, at a period of 7.7576 days.
We obtained 30 spectra of EBLM J0555-57A, using the high-resolution fibre-fed CORALIE \'echelle-spectrograph \citep{CORALIE_HARPS_QUELOZ}, mounted on the \textit{Euler} telescope, between 2013-11-14 and 2017-01-21. 

Two eclipse observations in the near-infrared $z'$-band were obtained with the \textit{Euler} \citep{Lendl_2013} and TRAPPIST \citep{Gillon_2011,Jehin_2011} telescopes, on the nights of 2014-02-24 and 2015-12-23 respectively. The observations reveal that our target was blended by a star that we label as J0555-57B. To confirm the source of the transit signal we compared observations with a large 38-pixel (px) aperture encompassing both stars, and a small 16 px aperture centred on the brighter star. A deeper transit signal was observed with the small aperture, identifying J0555-57A as the source of the eclipse signal. One spectrum of EBLM J0555-57B was obtained. The systemic radial velocities of the A and B components, $\gamma_A$ = 19.537 $\pm$ 0.015 kms$^{-1}$ and $\gamma_B$ = 19.968 $\pm$ 0.021 kms$^{-1}$, are nearly identical. A very similar position angle of the B component is observed on (blended) 2MASS images from 1999 and the \textit{Euler} image, which shows that A and B also share the same proper motion. This confirms that EBLM J0555-57A, B, and the transiting EBLM J0555-57Ab constitute a hierarchical triple system.
		
Focused images in the B, V, R, and $z'$-bands were obtained with \textit{Euler} on 2014-02-23 and 2016-01-10. We measured the separation between the primary and blend star to be $2.48 \pm 0.01''$, with a position angle, PA = $-105.57 \pm 0.23^\circ$. The magnitude difference, $\Delta z' =0.753\pm 0.035$ mag, translates into a flux-dilution of the eclipse depth by a factor $1.500 \pm 0.016$. The eclipse observations were reduced to obtain a photometric light-curve, as described in \citet{Lendl_2012} and \citet{Delrez_2014} for \textit{Euler} and TRAPPIST, respectively. 
A significant out-of-transit observation before ingress was obtained, but few out-of-transit measurements after egress could be made. Using literature broadband optical photometry and 2MASS J, H, and K magnitudes for the A and B components combined, together with the multi-colour magnitude differences ($\Delta b =0.95\pm 0.01$ mag, $\Delta r =0.786\pm 0.017$ mag, $\Delta v = 0.832\pm 0.014$ mag), we estimated IRFM temperatures \citep{Blackwell_1977} of $6450 \pm 200 K$ and $5950 \pm 200 K$ for the A and B components, respectively. A comparison to stellar model fluxes \citep{Kurucz_2004} was used, assuming a solar composition. The multi-colour observations and a \textit{GAIA} DR1 parallax measurement \citep{gaia_2016} were used to derive individual radii, R$_{\rm A} = 1.17 \pm 0.10 R_\odot$, and $R_{\rm B} = 0.94 \pm 0.08 R_\odot$. The parallax and angular separation of the A and B components determine a projected outer semi-major axis a$_{AB,p} = $ 479 $\pm$ 38 au.

\begin{figure}
	\centering
	\includegraphics[width= 0.24\textwidth]{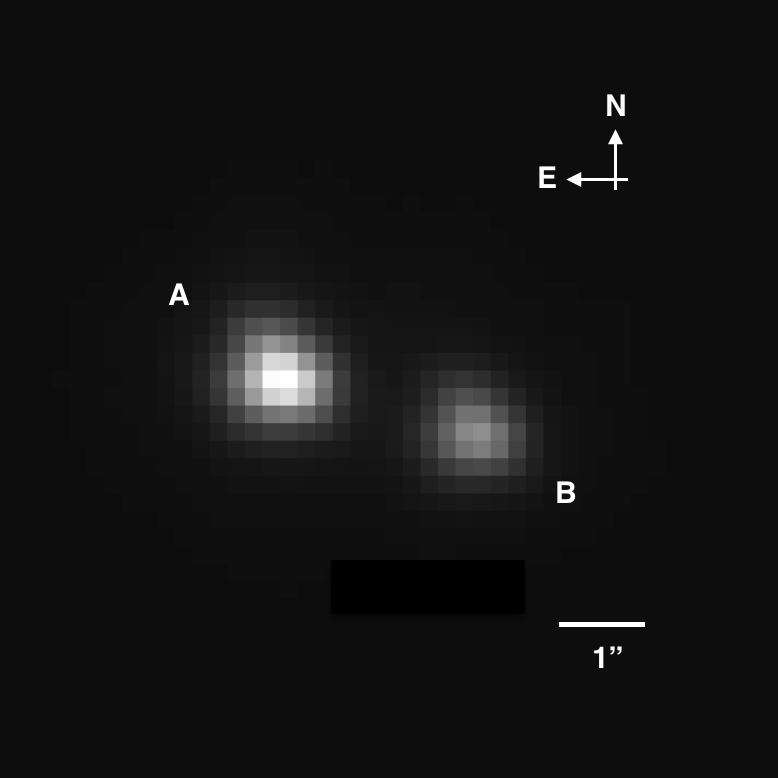}
	\caption{Focused image (z'-band) of EBLM~J0555-57 by \textit{Euler}, resolving the eclipsed (A) and tertiary (B) components.}
	\label{fig:ECAM}
\end{figure}

\begin{figure}
  	\centering
  	\includegraphics[width= 0.48\textwidth]{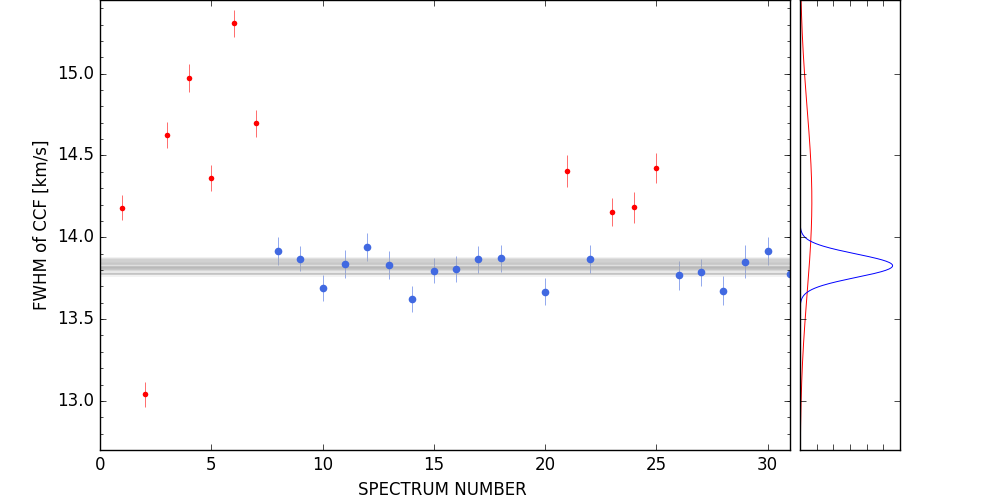}
  	\caption{FWHM of the CCF of the spectra of EBLM~J0555-57A. Blue: clean sample, red: contaminated sample. Grey: random draws from the posterior of the clean sample.}
  	\label{fig:FWHM}
\end{figure}

\section{Data analysis}

\subsection{Radial velocities} Radial velocities of EBLM J0555-57A were extracted by cross-correlating individual spectra with a numerical G2 mask \citep{Pepe_2002}. Varying seeing conditions resulted in fluctuations in the amount of flux from J0555-57B that enters the CORALIE fibre. This contamination can be identified by the full-width at half-maximum (FWHM) of the cross-correlation function (CCF). To select the non-contaminated spectra, we assumed two populations of points, a contaminated sample and a clean sample, with distinct means and variances. 
Following \citet{Hogg_2010}, a Markov chain Monte Carlo (MCMC) sampler was used to marginalise over the clean sample mean and variance, the contaminated sample mean and variance, and the prior probability that any point comes from the contaminated sample. We rejected a radial-velocity measurement and its associated spectrum when the FWHM had a posterior probability $<$ 1\% to originate from the clean distribution, as indicated in Fig.~\ref{fig:FWHM}. We excluded one point with a discrepant value in the span of the bisector inverse slope.

\subsection{Spectral Analysis} Atmospheric parameters were obtained via a wavelet-based Monte Carlo method  \citep{Gill2017}. The 18 spectra identified as uncontaminated were median-combined onto an identically sampled wavelength grid. After continuum regions were determined and normalised with spline functions, the spectrum was decomposed using a discrete Daubechies ($k = 4$) wavelet transform. We filtered out wavelet coefficients that corresponded to high-order noise and low-order systematics, associated with poor continuum placement. A grid of models was generated with the radiative transfer code {\sc SPECTRUM} \citep{Gray1994}, using {\sc MARCS} model atmospheres \citep{Gustafsson2008}, and version 5 of the GES atomic line list using {\sc{iSpec}} \citep{Blanco-Cuaresma_SPEC}, with solar abundances from \citet{Asplund2009}. Filtered coefficients were compared to those from the grid of models using an MCMC sampler implemented in {\sc emcee} \citep{emcee}. We used four free parameters, $\rm T_{\rm eff}$, [Fe/H], $\log g$, and $v\sin i_\star$, in the range 4000 -- 8000 K (250 K steps), l3.5 -- 5 dex (logg, 0.25 dex steps) and -1 -- 1 dex (Fe/H, 0.5 dex steps).

The median value of the cumulative posterior probability distribution was used to estimate the atmospheric parameters for J0555-57A (Table~\ref{tab:params2}). The precision associated with the wavelet method underestimates the uncertainty, so we adopt uncertainties by \citet{Blanco-CuaresmaSoubiranHeiterEtAl2014} for the synthetic spectral fitting technique of {\it GAIA} FGK benchmark stars ($124$~K, $0.21$~dex, and $0.14$~dex for $T_{\rm eff}$, $\log g$, and [Fe/H], respectively). The spectroscopic temperature measurements,T$_{\rm effA} = 6461 \pm 124$ K (18 spectra) and T$_{\rm effB} = 5717 \pm 124 $ K (1 spectrum) are consistent with the initial IRFM estimates.

\begin{figure}
	\centering
	\includegraphics[width= 0.48\textwidth]{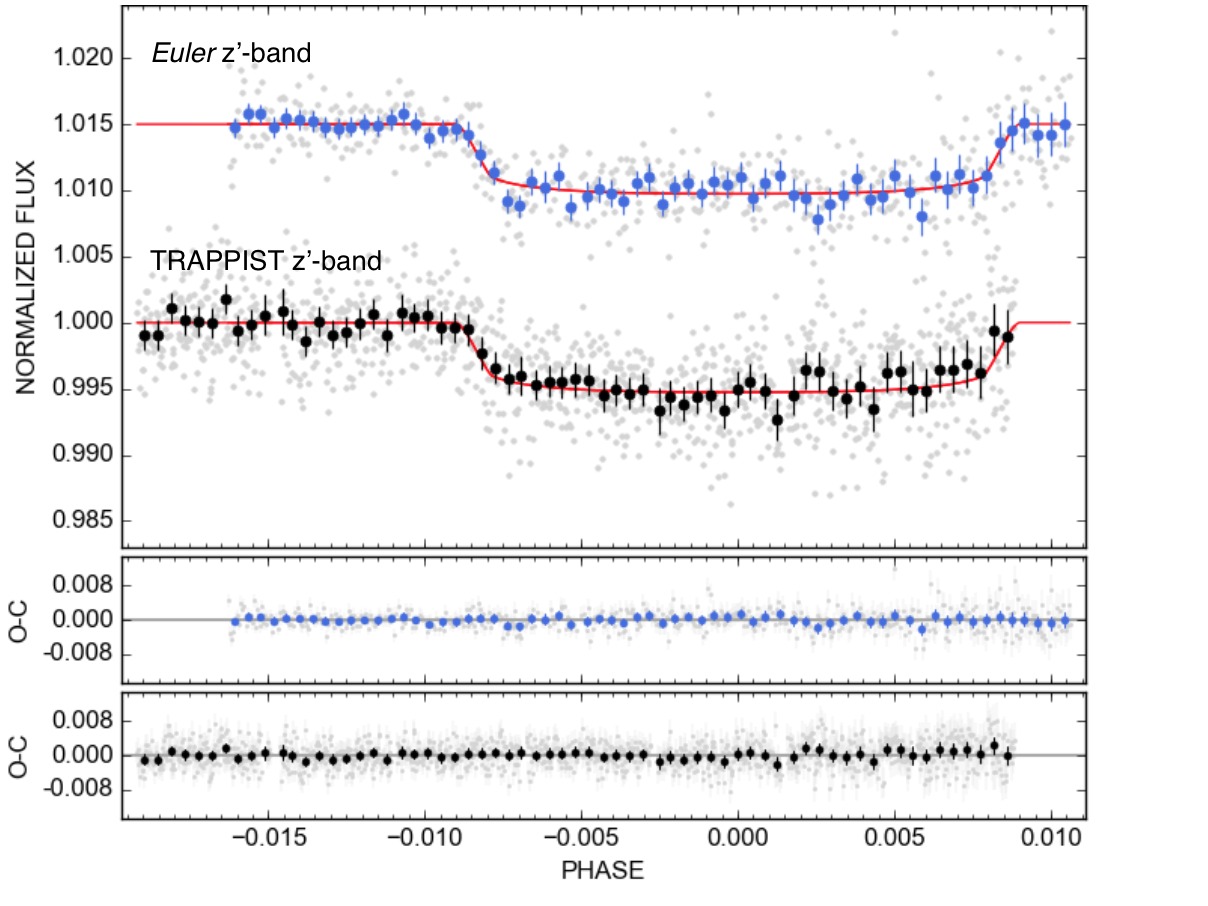}
	\caption{Transits of EBLM~J0555-57Ab, observed by \textit{Euler} (top), and TRAPPIST (bottom), with the best-fit model and residuals shown in the lower panels.}
	\label{fig:phot}
\end{figure}

\begin{figure}
	\includegraphics[width= 0.48\textwidth]{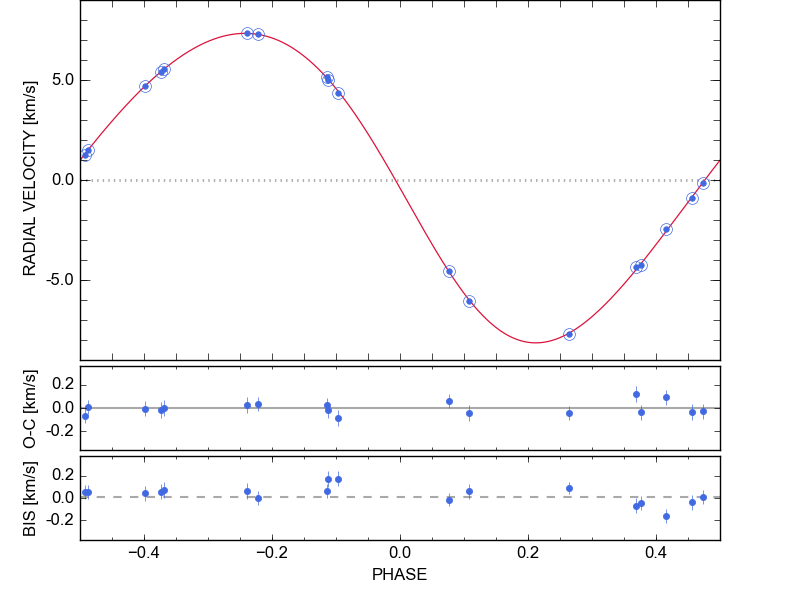}
	\caption{CORALIE radial velocities and Keplerian model for EBLM~J0555-57A. Uncertainties  are smaller than the symbols. Lower panels: residuals and span of bisector inverse slope (BIS).}
	\label{fig:rv}
\end{figure}

\section{Model of the data}

The radial velocity and light curves were modeled using the {\sc ellc} binary star model \citep{ellc}\footnote{We validated {\sc ellc} on two EBLM systems published in \citet{EBLM_1}, reaching a $1\sigma$-agreement on the derived parameters.}. 
An MCMC sampler \citep[{\sc emcee};][]{emcee}, was used to fit the transit light-curves and radial velocities in a framework similar to that described in \citet{EBLM_1}. 
 We used the Bayesian information criterion (BIC) \citep{Schwarz_1978} to compare detrending baselines of varying complexity in time-, position-, FWHM-, and background-dependence. 
 A flat baseline with a linear background subtraction is preferred for both light curves. Indeed both observations show a large fluctuation in the background flux. 

The parameters used in the MCMC sampling are the period $P$, the mid-transit time $t_0$, the observed transit depth $D_{\rm obs}$, the transit duration $W$, the impact parameter $b$, the semi-amplitude $K$, the parameters $\sqrt{e} \sin \omega$, $\sqrt{e} \cos \omega$, and the systemic velocity $\gamma$. The RV sample was separated into two parts, with distinct systemic velocities, to account for a change in the zero-point of CORALIE after a recent upgrade \citep{triaud_2017}.
The geometric parameters of the system, $R_1 / a$, $R_2 / a$,  and $i$ were derived from the MCMC parameters using the formalism from \citet{Winn_2010}, and were then passed to {\sc ellc}. 
The TRAPPIST sequence was interrupted by a meridian flip; to account for a possible systematic offset in the flux measurement, an offset-factor was included for measurements before the flip.\\
We used a quadratic limb-darkening law in the MCMC analysis, with a Gaussian prior on coefficients that were interpolated from \citet{Claret_2004}, using the spectroscopic parameters $\rm T_{\rm eff}$, [Fe/H] and $\log g$. We included nuisance parameters in the MCMC sampler, that scale uncertainties in the photometry and radial velocity to account for white noise. The baseline parameters for a linear background subtraction, meridian flip, and normalization are fitted by a least-squares algorithm. 
Where not explicitly stated otherwise, we used unbounded, or sensibly bounded uniform priors to constrain parameters to physical intervals, for instance (0~<~e~<~1). The B-component dilutes the transit depth by a factor $f_{d} = {f_{\rm A}}/({f_{\rm A} + f_{\rm B}})$, where $f_A$ and $f_B$ denote the flux from the A and B components respectively. We sampled a Gaussian prior on this depth dilution factor, $f_{d} = 1.500 \pm 0.016$, to compute the true transit depth $D_{\rm calc}$ at every step in the Markov chains. This calculated transit depth was used in the derivation of the physical parameters.

We analyzed this first global fit for correlated noise in the photometry \citep{Gillon_2012,Winn_2008}. The light curve was binned in the range of 10 to 30 min and the maximum root-mean square (RMS) deviation of the residuals in this bin range was determined. The flux uncertainties were then rescaled by the ratio of the maximum binned RMS deviation to the RMS deviation of the un-binned residuals. This increased the uncertainties by factors of 2.02 and 1.37 for TRAPPIST and \textit{Euler} respectively. We then performed a global MCMC fit using 100 chains of 10 000 steps each.

The modes of the marginalised posterior distributions for each jump parameter are reported with upper and lower 68\% confidence intervals. 
The physical parameters of the system were derived from the MCMC parameters, in particular the parameter $\log g_2$, which is independent of the primary star mass \citep{Southworth_2004}.
We used the primary star density to iteratively refine the primary mass estimate M$_1$. An initial primary density was estimated from the transit and was used to determine a primary mass using {\sc bagemass} \citep{bagemass}. {\sc bagemass} uses stellar evolution models by \citet{Weiss_2007}. The primary star mass was then used with the transit and radial velocity model, to compute an updated density, and we proceeded iteratively. The calculated density was found to be consistent from the first iteration step. 

\section{Results}

Independently of any assumptions for the primary star, we obtain a surface gravity $\log g_2 =  5.50^{+0.03}_{-0.13}$ for EBLM~J0555-57Ab, comparable to that of the recently announced brown dwarf EPIC~201702477b \citep{Bayliss_2017}.  We determine a mass function f(m) = $0.0003686\,^{+0.0000037}_{-0.0000049}$ M$_\odot$. Using the primary star mass determined with {\sc bagemass}, we find a stellar companion with mass $85.2^{+4.0}_{-3.9}$ M$_{\rm Jup}$ ($0.0813^{+0.0038}_{-0.0037}$ M$_{\odot}$) and radius $0.84^{+0.14}_{-0.04}$ R$_{\rm Jup}$ ($0.084^{+0.014}_{-0.004}$ R$_{\odot}$). This implies a mass ratio $q$ = 0.0721$^{+0.0019}_{-0.0017}$. A lower uncertainty in the radius measurement may be achievable by high-precision photometry \citep[e.g. TESS;][]{Sullivan_2015}. The fit of the radial velocity results in an  RMS deviation of 65 ms$^{-1}$, and our analysis reveals a low but significant orbital eccentricity, $e = 0.0894^{+0.0035}_{-0.0036}$. The BIC of a forced circular fit, and the Lucy-Sweeney test \citep{Lucy_1971} validate this orbital eccentricity, since its measurement is significant at $\sim$25$\sigma$. The non-zero eccentricity of EBLM J0555-57Ab could indicate a previous orbital decay, for instance by Kozai-Lidov oscillations \citep{Lidov_1961,Kozai_1962} induced by J0555-57B, or an undetected body, followed by tidal friction \citep{Fabrycky_2007}. At the current semi-major axis, a $=$ 0.0817 au, such Kozai-Lidov oscillations are likely suppressed by general-relativistic precession \citep{Fabrycky_2007,Cristobal_2014}. It is unlikely that a contamination of the spectra causes the measured non-zero eccentricity, but further spectroscopic observations with a fibre of smaller diameter can clarify this. We note a discrepancy between the spectroscopic log$g_{\rm 1spec} = 4.18 \pm 0.21$ and that derived from the calculated radius and prior mass, log$g_1 = 4.5^{+0.03}_{-0.13}$. Spectroscopic measurements of log$g$ are known to be poorly constrained \citep{Torres_2012, Bruntt_2012, Doyle_2015}. We verify that adopting a prior on log$g_1$ for the spectroscopic analysis, using the derived value, leads to a primary and companion mass and radius that are consistent with the previous result.

\begin{figure}
	\includegraphics[width=\hsize]{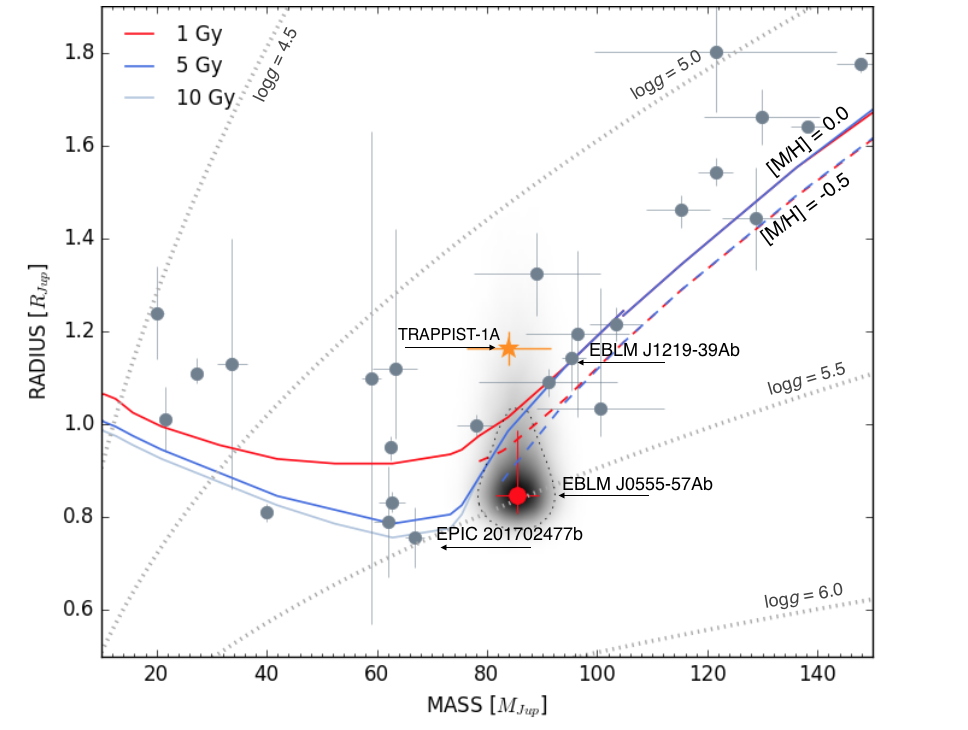}
	\caption{Mass-radius posterior distribution for EBLM~J0555-57Ab with an 68\% confidence region (dashed). Isochrones for solar metalicity \citep{Baraffe_2015}, and sub-solar metalicity [M/H = -0.5] \citep{Baraffe_98} are plotted. Objects from  \citet{EBLM_1, S_gransan_2003, Demory_2009, Bayliss_2017, D_az_2014, Johnson_2011, Siverd_2012} and \citet{Chen_2014} are also shown.} 
	\label{fig:massradius}%
\end{figure}

We conclude that EBLM~J0555-57Ab is located just above the hydrogen-burning mass limit that separates stellar and sub-stellar objects ($\sim83$ M$_{\rm Jup}$ for objects with [M/H] = -0.5; \citet{Baraffe_98}). In Figure~\ref{fig:massradius} we show the posterior distribution of J0555-57Ab on the mass-radius diagram for brown dwarfs and low-mass stars. Our results using {\sc bagemass} indicate an age of 1.9 $\pm$ 1.2 Gy for J0555-57A. The mass and radius of J0555-57Ab are consistent with models of a metal-poor, low-mass star. J0555-57Ab does not show evidence of a radius that is inflated, for instance by magnetic fields, as hypothesized by \citet{Lopez_Morales_2007} for low-mass stars. With its location on the lower bound of the mass-radius relation for stellar objects, J0555-57Ab is a critical object in the empirical calibration of the mass-radius relation in this regime. J0555-57Ab has a mass similar to that of TRAPPIST-1A. \citep{Gillon_2016,Gillon_2017}. The low radius of EBLM~J0555-57Ab, comparable to that of the low-mass star 2MASS~J0523-1403 \citep{Dieterich_2014}, demonstrates the size dispersion for low-mass stars. It is essential that such variations are understood as we prepare for the detection of multi-planetary systems orbiting ultra-cool dwarfs by experiments such as SPECULOOS \citep{Gillon_2013}.  

\begin{table}
\caption{Parameters of {EBLM~J0555-57A and Ab}. The last two significant figures of the uncertainty are shown in brackets. Dates are BJD$_{\rm UTC}$ - 2 450 000. R$_{\rm Jup} =  69.911 \cdot 10^6$ km.}

\renewcommand{\arraystretch}{1.2} 

\begin{tabular}{llll} \hline \hline
\multicolumn{4}{c}{1SWASP J055532.69-571726.0}\\
\multicolumn{4}{c}{2MASS J05553262-5717261}\\
\multicolumn{4}{c}{TYC 8528-926-1; CD-57 1311 }\\
\hline

Parameter  &  & Parameter  &  \\ \hline

\multicolumn{4}{l}{\textit{Spectral line analysis of primary; system parameters}}\\

$T_{\rm eff}$    	 	 	& $6461 \pm 124$ K			        &{[Fe/H]}  		  			 &$-0.24 \pm 0.16 $ \\
$\log g_\mathrm{1}$	  & $4.18 \pm 0.21$ (cgs) 			  &$v \sin i_1$	  			 & $7.60\pm 0.28$ km/s \\
d\tablefootmark{*} & 193.8$^{+14.5}_{-12.6}$ pc & Mag \tablefootmark{**} & A: 9.98, B: 10.76 \\
\hline 
 \multicolumn{4}{l}{\textit{Parameters of MCMC}}\\
 $P$							&	7.757676$^{(+29)}_{(-25)}$ d & $t_{0}$ &	6\,712.6452 $^{(+15)}_{(-14)}$\\
 $D_{\rm obs}$			 &	0.00474 $^{(+26)}_{(-21)}$		& $W$			  			 &	0.1392 $^{(+49)}_{(-28)}$ d\\
 $\sqrt{e} \cos \omega$&	$-$0.176 $^{(+06)}_{(-06)}$        &$b$						&	0.35\tablefootmark{***} $^{+0.24}_{-0.25}$ R$_\odot$	\\
$\sqrt{e} \sin \omega$	&	0.241 $^{(+09)}_{(-09)}$		& $K_\mathrm{1}$	&	7.740 $^{(+27)}_{(-33)}$ km\,s$^{-1}$\\
 \hline
 \multicolumn{4}{l}{\textit{Derived parameters from MCMC}}\\
$D_{\rm calc}$				& $0.00713^{(+41)}_{(-33)}$			&$f(m)$						&	$0.0003686\,^{(+37)}_{(-49)}$ M$_\odot$\\
$R_\mathrm{2}$/$R_\mathrm{1}$	& $0.0844^{(+23)}_{(-19)}$&$\log\,g_\mathrm{2}$	&	5.50 $^{+0.03}_{-0.13}$ (cgs)\\
$R_\mathrm{1}$/a				&0.0562 $^{(+88)}_{(-13)}	$				&$R_\mathrm{2}$/a		&	0.00480 $^{(+79)}_{(-16)}$\\
$M_\mathrm{1}$	(prior)		&$1.13 \pm 0.08$ $M_\odot$				   &$M_\mathrm{2}$		&	85.2 $^{+4.0}_{-3.9}$ $M_\mathrm{Jup}$\\
$R_\mathrm{1}$				&0.99 $^{+0.15}_{-0.03}$ $R_\odot$		&$R_\mathrm{2}$		&	0.84 $^{+0.14}_{-0.04}$ $R_\mathrm{Jup}$\\
$\rho_1$ & $1.16^{+0.1}_{-0.42}$ $\rho_\odot$	& $\rho_2$ & $188^{+25}_{-69}$ g\,cm$^{-3}$ \\ 
$a$							&$0.0817^{(+19)}_{(-19)}$ au		  &$i$					&	89.84 $^{+0.2}_{-1.8}$ deg\\
$e$							& $0.0894^{(+35)}_{(-36)}$				&$\omega $		&	$-53.7^{+1.5}_{-1.8}$ deg\\ 
$q$	& 0.0721 $^{(+19)}_{(-17)}$ \\
\hline
\end{tabular}
\tablefoot{\tablefoottext{*}{from GAIA parallax}
	\tablefoottext{**}{GAIA g-magnitude}
	\tablefoottext{***}{The mode is ambiguous so the median value is provided.}}
\label{tab:params2}
\end{table}

\begin{acknowledgements}
We thank the anonymous referee for valuable comments that improved the manuscript. The Swiss {\it Euler} Telescope is funded by the Swiss National Science Foundation. TRAPPIST-South is a project funded by the Belgian Fonds (National) de la Recherche Scientifique (F.R.S.-FNRS) under grant FRFC 2.5.594.09.F, with the participation of the Swiss National Science Foundation (FNS/SNSF). WASP-South is hosted by the South African Astronomical Observatory and we are grateful for their ongoing support and assistance. L. Delrez acknowledges support from the Gruber Foundation Fellowship. M. Gillon and E. Jehin are Belgian F.R.S.-FNRS Research Associates. This work was partially supported by a grant from the Simons Foundation (PI Queloz, grant number 327127). 

\end{acknowledgements}

\bibliographystyle{aa}
\bibliography{j0555references.bib}

\begin{thebibliography}{61}
\expandafter\ifx\csname natexlab\endcsname\relax\def\natexlab#1{#1}\fi

\bibitem[{{Asplund} {et~al.}(2009){Asplund}, {Grevesse}, {Sauval}, \&
  {Scott}}]{Asplund2009}
{Asplund}, M., {Grevesse}, N., {Sauval}, A.~J., \& {Scott}, P. 2009, \araa, 47,
  481

\bibitem[{Baraffe {et~al.}(1998)Baraffe, Chabrier, Allard, \&
  Hauschildt}]{Baraffe_98}
Baraffe, I., Chabrier, G., Allard, F., \& Hauschildt, P. 1998, A\&A

\bibitem[{Baraffe {et~al.}(2015)Baraffe, Homeier, Allard, \&
  Chabrier}]{Baraffe_2015}
Baraffe, I., Homeier, D., Allard, F., \& Chabrier, G. 2015, A\&A, 577, A42

\bibitem[{Bayliss {et~al.}(2017)Bayliss, Hojjatpanah, Santerne, Dragomir, Zhou,
  Shporer, Col{\'{o}}n, Almenara, Armstrong, Barrado, Barros, Bento, Boisse,
  Bouchy, Brown, Brown, Cameron, Cochran, Demangeon, Deleuil, D{\'{\i}}az,
  Fulton, Horne, H{\'{e}}brard, Lillo-Box, Lovis, Mawet, Ngo, Osborn, Palle,
  Petigura, Pollacco, Santos, Sefako, Siverd, Sousa, \&
  Tsantaki}]{Bayliss_2017}
Bayliss, D., Hojjatpanah, S., Santerne, A., {et~al.} 2017, \AJ, 153, 15

\bibitem[{Berta-Thompson {et~al.}(2015)Berta-Thompson, Irwin, Charbonneau,
  Newton, Dittmann, Astudillo-Defru, Bonfils, Gillon, Jehin, Stark, Stalder,
  Bouchy, Delfosse, Forveille, Lovis, Mayor, Neves, Pepe, Santos, Udry, \&
  Wünsche}]{Zachory_2015}
Berta-Thompson, Z.~K., Irwin, J., Charbonneau, D., {et~al.} 2015, Nature

\bibitem[{Blackwell \& Shallis(1977)}]{Blackwell_1977}
Blackwell, D.~E. \& Shallis, M.~J. 1977, MNRAS, 180, 177

\bibitem[{{Blanco-Cuaresma} {et~al.}(2016){Blanco-Cuaresma}, {Nordlander},
  {Heiter}, {Jofr{\'e}}, {Masseron}, {Casamiquela}, {Tabernero}, {Bhat},
  {Casey}, {Mel{\'e}ndez}, \& {Ram{\'{\i}}rez}}]{Blanco-Cuaresma_SPEC}
{Blanco-Cuaresma}, S., {Nordlander}, T., {Heiter}, U., {et~al.} 2016, in 19th
  Cambridge Workshop on Cool Stars, Stellar Systems, and the Sun (CS19), 22

\bibitem[{{Blanco-Cuaresma} {et~al.}(2014){Blanco-Cuaresma}, {Soubiran},
  {Heiter}, \& {Jofr{\'e}}}]{Blanco-CuaresmaSoubiranHeiterEtAl2014}
{Blanco-Cuaresma}, S., {Soubiran}, C., {Heiter}, U., \& {Jofr{\'e}}, P. 2014,
  \aap, 569, A111

\bibitem[{Bruntt {et~al.}(2012)Bruntt, Basu, Smalley, Chaplin, Verner, Bedding,
  Catala, Gazzano, Molenda-{\.{Z}}akowicz, Thygesen, Uytterhoeven, Hekker,
  Huber, Karoff, Mathur, Mosser, Appourchaux, Campante, Elsworth,
  Garc{\'{\i}}a, Handberg, Metcalfe, Quirion, R{\'{e}}gulo, Roxburgh, Stello,
  Christensen-Dalsgaard, Kawaler, Kjeldsen, Morris, Quintana, \&
  Sanderfer}]{Bruntt_2012}
Bruntt, H., Basu, S., Smalley, B., {et~al.} 2012, MNRAS, 423, 122

\bibitem[{Castelli \& Kurucz(2004)}]{Kurucz_2004}
Castelli, F. \& Kurucz, R.~L. 2004, ArXiv e-prints, arXiv:astro-ph/0405087v1

\bibitem[{Chabrier(2003)}]{Chabrier_2003}
Chabrier, G. 2003, PASP, 115, 763

\bibitem[{Chen {et~al.}(2014)Chen, Girardi, Bressan, Marigo, Barbieri, \&
  Kong}]{Chen_2014}
Chen, Y., Girardi, L., Bressan, A., {et~al.} 2014, MNRAS, 444, 2525

\bibitem[{Claret(2004)}]{Claret_2004}
Claret, A. 2004, A\&A, 428, 1001

\bibitem[{{Collier Cameron} {et~al.}(2007){Collier Cameron}, Wilson, West,
  Hebb, Wang, Aigrain, Bouchy, Christian, Clarkson, Enoch, Esposito, Guenther,
  Haswell, H{\'{e}}brard, Hellier, Horne, Irwin, Kane, Loeillet, Lister,
  Maxted, Mayor, Moutou, Parley, Pollacco, Pont, Queloz, Ryans, Skillen,
  Street, Udry, \& Wheatley}]{Collier_Cameron_2007}
{Collier Cameron}, A., Wilson, D.~M., West, R.~G., {et~al.} 2007, MNRAS, 380,
  1230

\bibitem[{{de Wit} {et~al.}(2016){de Wit}, {Wakeford}, {Gillon}, {Lewis},
  {Valenti}, {Demory}, {Burgasser}, {Burdanov}, {Delrez}, {Jehin}, {Lederer},
  {Queloz}, {Triaud}, \& {Van Grootel}}]{deWit_2016}
{de Wit}, J., {Wakeford}, H.~R., {Gillon}, M., {et~al.} 2016, \nat, 537, 69

\bibitem[{{Delrez} {et~al.}(2014){Delrez}, {Van Grootel}, {Anderson},
  {Collier-Cameron}, {Doyle}, {Fumel}, {Gillon}, {Hellier}, {Jehin}, {Lendl},
  {Neveu-VanMalle}, {Maxted}, {Pepe}, {Pollacco}, {Queloz}, {S{\'e}gransan},
  {Smalley}, {Smith}, {Southworth}, {Triaud}, {Udry}, \& {West}}]{Delrez_2014}
{Delrez}, L., {Van Grootel}, V., {Anderson}, D.~R., {et~al.} 2014, \aap, 563,
  A143

\bibitem[{Demory {et~al.}(2009)Demory, S{\'{e}}gransan, Forveille, Queloz,
  Beuzit, Delfosse, Folco, Kervella, Bouquin, Perrier, Benisty, Duvert,
  Hofmann, Lopez, \& Petrov}]{Demory_2009}
Demory, B.-O., S{\'{e}}gransan, D., Forveille, T., {et~al.} 2009, A\&A, 505,
  205

\bibitem[{D{\'{\i}}az {et~al.}(2014)D{\'{\i}}az, Montagnier, Leconte, Bonomo,
  Deleuil, Almenara, Barros, Bouchy, Bruno, Damiani, H{\'{e}}brard, Moutou, \&
  Santerne}]{D_az_2014}
D{\'{\i}}az, R.~F., Montagnier, G., Leconte, J., {et~al.} 2014, A\&A, 572, A109

\bibitem[{Dieterich {et~al.}(2014)Dieterich, Henry, Jao, Winters, Hosey,
  Riedel, \& Subasavage}]{Dieterich_2014}
Dieterich, S.~B., Henry, T.~J., Jao, W.-C., {et~al.} 2014, \AJ, 147, 94

\bibitem[{Doyle(2015)}]{Doyle_2015}
Doyle, A.~P. 2015, PhD thesis, Keele University

\bibitem[{Fabrycky \& Tremaine(2007)}]{Fabrycky_2007}
Fabrycky, D. \& Tremaine, S. 2007, ApJ

\bibitem[{{Foreman-Mackey} {et~al.}(2013){Foreman-Mackey}, {Hogg}, {Lang}, \&
  {Goodman}}]{emcee}
{Foreman-Mackey}, D., {Hogg}, D.~W., {Lang}, D., \& {Goodman}, J. 2013, \pasp,
  125, 306

\bibitem[{{Gaia Collaboration} {et~al.}(2016){Gaia Collaboration}, {Brown},
  {Vallenari}, {Prusti}, {de Bruijne}, {Mignard}, {Drimmel}, {Babusiaux},
  {Bailer-Jones}, {Bastian}, \& et~al.}]{gaia_2016}
{Gaia Collaboration}, {Brown}, A.~G.~A., {Vallenari}, A., {et~al.} 2016, \aap,
  595, A2

\bibitem[{Gill {et~al.}(2017)Gill, Maxted, \& Smalley}]{Gill2017}
Gill, S., Maxted, P. F.~L., \& Smalley, B. 2017, in prep

\bibitem[{{Gillon} {et~al.}(2013){Gillon}, {Jehin}, {Delrez}, {Magain},
  {Opitom}, \& {Sohy}}]{Gillon_2013}
{Gillon}, M., {Jehin}, E., {Delrez}, L., {et~al.} 2013, in Protostars and
  Planets VI Posters

\bibitem[{Gillon {et~al.}(2016)Gillon, Jehin, Lederer, Delrez, de~Wit,
  Burdanov, Grootel, Burgasser, Triaud, Opitom, Demory, Sahu, Gagliuffi,
  Magain, \& Queloz}]{Gillon_2016}
Gillon, M., Jehin, E., Lederer, S.~M., {et~al.} 2016, Nature, 533, 221

\bibitem[{Gillon {et~al.}(2011)Gillon, Jehin, Magain, Chantry,
  Hutsem{\'{e}}kers, Manfroid, Queloz, \& Udry}]{Gillon_2011}
Gillon, M., Jehin, E., Magain, P., {et~al.} 2011, {EPJ} Web of Conferences, 11,
  06002

\bibitem[{{Gillon} {et~al.}(2017){Gillon}, {Triaud}, {Demory}, {Jehin}, {Agol},
  {Deck}, {Lederer}, {de Wit}, {Burdanov}, {Ingalls}, {Bolmont}, {Leconte},
  {Raymond}, {Selsis}, {Turbet}, {Barkaoui}, {Burgasser}, {Burleigh}, {Carey},
  {Chaushev}, {Copperwheat}, {Delrez}, {Fernandes}, {Holdsworth}, {Kotze}, {Van
  Grootel}, {Almleaky}, {Benkhaldoun}, {Magain}, \& {Queloz}}]{Gillon_2017}
{Gillon}, M., {Triaud}, A.~H.~M.~J., {Demory}, B.-O., {et~al.} 2017, \nat, 542,
  456

\bibitem[{Gillon {et~al.}(2012)Gillon, Triaud, Fortney, Demory, Jehin, Lendl,
  Magain, Kabath, Queloz, Alonso, Anderson, Cameron, Fumel, Hebb, Hellier,
  Lanotte, Maxted, Mowlavi, \& Smalley}]{Gillon_2012}
Gillon, M., Triaud, A. H. M.~J., Fortney, J.~J., {et~al.} 2012, A\&A, 542, A4

\bibitem[{{Gray} \& {Corbally}(1994)}]{Gray1994}
{Gray}, R.~O. \& {Corbally}, C.~J. 1994, \aj, 107, 742

\bibitem[{{Gustafsson} {et~al.}(2008){Gustafsson}, {Edvardsson}, {Eriksson},
  {J{\o}rgensen}, {Nordlund}, \& {Plez}}]{Gustafsson2008}
{Gustafsson}, B., {Edvardsson}, B., {Eriksson}, K., {et~al.} 2008, \aap, 486,
  951

\bibitem[{{Gómez Maqueo Chew} {et~al.}(2014){Gómez Maqueo Chew}, Morales,
  Faedi, García-Melendo, Hebb, Rodler, Deshpande, Mahadevan, McCormac, Barnes,
  Triaud, López-Morales, Skillen, Cameron, Joner, Laney, Stephens, Stassun, \&
  Montañés-Rodríguez}]{EBLM_II}
{Gómez Maqueo Chew}, Y., Morales, J.~C., Faedi, F., {et~al.} 2014, \aap

\bibitem[{Henry {et~al.}(2006)Henry, Jao, Subasavage, Beaulieu, Ianna, Costa,
  \& M{\'{e}}ndez}]{Henry_2006}
Henry, T.~J., Jao, W.-C., Subasavage, J.~P., {et~al.} 2006, ApJ, 132

\bibitem[{{Hogg} {et~al.}(2010){Hogg}, {Bovy}, \& {Lang}}]{Hogg_2010}
{Hogg}, D.~W., {Bovy}, J., \& {Lang}, D. 2010, ArXiv e-prints,
  arXiv:1008.4686v1

\bibitem[{Jehin {et~al.}(2011)Jehin, Gillon, Queloz, Magain, Manfroid, Chandry,
  Lendl, Hutsemekers, \& Udry}]{Jehin_2011}
Jehin, E., Gillon, M., Queloz, D., {et~al.} 2011, The Messenger, 145

\bibitem[{Johnson {et~al.}(2011)Johnson, Apps, Gazak, Crepp, Crossfield,
  Howard, Marcy, Morton, Chubak, \& Isaacson}]{Johnson_2011}
Johnson, J.~A., Apps, K., Gazak, J.~Z., {et~al.} 2011, ApJ, 730, 79

\bibitem[{Kozai(1962)}]{Kozai_1962}
Kozai, Y. 1962, \aj, 67, 591

\bibitem[{{Kroupa}(2001)}]{Kroupa_2001}
{Kroupa}, P. 2001, \mnras, 322, 231

\bibitem[{{Lendl} {et~al.}(2012){Lendl}, {Anderson}, {Collier-Cameron},
  {Doyle}, {Gillon}, {Hellier}, {Jehin}, {Lister}, {Maxted}, {Pepe},
  {Pollacco}, {Queloz}, {Smalley}, {S{\'e}gransan}, {Smith}, {Triaud}, {Udry},
  {West}, \& {Wheatley}}]{Lendl_2012}
{Lendl}, M., {Anderson}, D.~R., {Collier-Cameron}, A., {et~al.} 2012, \aap,
  544, A72

\bibitem[{Lendl {et~al.}(2013)Lendl, Gillon, Queloz, Alonso, Fumel, Jehin, \&
  Naef}]{Lendl_2013}
Lendl, M., Gillon, M., Queloz, D., {et~al.} 2013, \aap, 552, A2

\bibitem[{Lidov(1961)}]{Lidov_1961}
Lidov, M.~L. 1961, Iskusst. Sputniki Zemli, 8, 5

\bibitem[{Lopez-Morales(2007)}]{Lopez_Morales_2007}
Lopez-Morales, M. 2007, ApJ, 660, 732

\bibitem[{Lucy \& Sweeney(1971)}]{Lucy_1971}
Lucy, L.~B. \& Sweeney, M.~A. 1971, \AJ, 76, 544

\bibitem[{{Luger} {et~al.}(2017){Luger}, {Sestovic}, {Kruse}, {Grimm},
  {Demory}, {Agol}, {Bolmont}, {Fabrycky}, {Fernandes}, {Van Grootel},
  {Burgasser}, {Gillon}, {Ingalls}, {Jehin}, {Raymond}, {Selsis}, {Triaud},
  {Barclay}, {Barentsen}, {Delrez}, {de Wit}, {Foreman-Mackey}, {Holdsworth},
  {Leconte}, {Lederer}, {Turbet}, {Almleaky}, {Benkhaldoun}, {Magain},
  {Morris}, {Heng}, \& {Queloz}}]{Luger_2017}
{Luger}, R., {Sestovic}, M., {Kruse}, E., {et~al.} 2017, Nature Astronomy, 1

\bibitem[{Maxted(2016)}]{ellc}
Maxted, P. F.~L. 2016, A\&A 591, A111

\bibitem[{Maxted {et~al.}(2014)Maxted, Serenelli, \& Southworth}]{bagemass}
Maxted, P. F.~L., Serenelli, A.~M., \& Southworth, J. 2014, A\&A 575

\bibitem[{{Pepe} {et~al.}(2002){Pepe}, {Mayor}, {Rupprecht}, {Avila},
  {Ballester}, {Beckers}, {Benz}, {Bertaux}, {Bouchy}, {Buzzoni}, {Cavadore},
  {Deiries}, {Dekker}, {Delabre}, {D'Odorico}, {Eckert}, {Fischer}, {Fleury},
  {George}, {Gilliotte}, {Gojak}, {Guzman}, {Koch}, {Kohler}, {Kotzlowski},
  {Lacroix}, {Le Merrer}, {Lizon}, {Lo Curto}, {Longinotti}, {Megevand},
  {Pasquini}, {Petitpas}, {Pichard}, {Queloz}, {Reyes}, {Richaud}, {Sivan},
  {Sosnowska}, {Soto}, {Udry}, {Ureta}, {van Kesteren}, {Weber}, {Weilenmann},
  {Wicenec}, {Wieland}, {Christensen-Dalsgaard}, {Dravins}, {Hatzes},
  {K{\"u}rster}, {Paresce}, \& {Penny}}]{Pepe_2002}
{Pepe}, F., {Mayor}, M., {Rupprecht}, G., {et~al.} 2002, The Messenger, 110, 9

\bibitem[{Petrovich(2014)}]{Cristobal_2014}
Petrovich, C. 2014, ApJ, 779

\bibitem[{Pollacco {et~al.}(2006)Pollacco, Skillen, Cameron, Christian,
  Hellier, Irwin, Lister, Street, West, Anderson, Clarkson, Deeg, Enoch, Evans,
  Fitzsimmons, Haswell, Hodgkin, Horne, Kane, Keenan, Maxted, Norton, Osborne,
  Parley, Ryans, Smalley, Wheatley, \& Wilson}]{Pollacco_2006}
Pollacco, D.~L., Skillen, I., Cameron, A.~C., {et~al.} 2006, PASP, 118, 1407

\bibitem[{Queloz {et~al.}(2001)Queloz, Mayor, Udry, Burnet, Carrier,
  Eggenberger, Naef, Santos, Pepe, Rupprecht, Avila, Baeza, Benz, Bertaux,
  Bouchy, Cavadore, Delabre, Eckert, Fischer, Fleury, Gilliotte, Goyak, Guzman,
  Kohler, Lacroix, Lizon, Megevand, Sivan, Sosnowska, \&
  Weilenmann}]{CORALIE_HARPS_QUELOZ}
Queloz, D., Mayor, M., Udry, S., {et~al.} 2001, The Messenger, 105, 1

\bibitem[{Schwarz(1978)}]{Schwarz_1978}
Schwarz, G. 1978, The Annals of Statistics, 6, 461

\bibitem[{S{\'{e}}gransan {et~al.}(2003)S{\'{e}}gransan, Kervella, Forveille,
  \& Queloz}]{S_gransan_2003}
S{\'{e}}gransan, D., Kervella, P., Forveille, T., \& Queloz, D. 2003, A\&A,
  397, L5

\bibitem[{Siverd {et~al.}(2012)Siverd, Beatty, Pepper, Eastman, Collins,
  Bieryla, Latham, Buchhave, Jensen, Crepp, Street, Stassun, Gaudi, Berlind,
  Calkins, DePoy, Esquerdo, Fulton, Furesz, Geary, Gould, Hebb, Kielkopf,
  Marshall, Pogge, Stanek, Stefanik, Szentgyorgyi, Trueblood, Trueblood, Stutz,
  \& van Saders}]{Siverd_2012}
Siverd, R.~J., Beatty, T.~G., Pepper, J., {et~al.} 2012, ApJ

\bibitem[{Southworth {et~al.}(2004)Southworth, Zucker, Maxted, \&
  Smalley}]{Southworth_2004}
Southworth, J., Zucker, S., Maxted, P. F.~L., \& Smalley, B. 2004, MNRAS, 355,
  986

\bibitem[{{Sullivan} {et~al.}(2015){Sullivan}, {Winn}, {Berta-Thompson},
  {Charbonneau}, {Deming}, {Dressing}, {Latham}, {Levine}, {McCullough},
  {Morton}, {Ricker}, {Vanderspek}, \& {Woods}}]{Sullivan_2015}
{Sullivan}, P.~W., {Winn}, J.~N., {Berta-Thompson}, Z.~K., {et~al.} 2015, \apj,
  809, 77

\bibitem[{Torres {et~al.}(2012)Torres, Fischer, Sozzetti, Buchhave, Winn,
  Holman, \& Carter}]{Torres_2012}
Torres, G., Fischer, D.~A., Sozzetti, A., {et~al.} 2012, \apj, 757, 161

\bibitem[{Triaud {et~al.}(2013)Triaud, Hebb, Anderson, Cargile,
  Collier~Cameron, Doyle, Faedi, Gillon, Gomez Maqueo~Chew, Hellier, Jehin,
  Maxted, Naef, Pepe, Pollacco, Queloz, S{\'e}gransan, Smalley, Stassun, Udry,
  \& West}]{EBLM_1}
Triaud, A. H. M.~J., Hebb, L., Anderson, D.~R., {et~al.} 2013, \aap, 549, A18

\bibitem[{{Triaud} {et~al.}(2017){Triaud}, {Neveu-VanMalle}, {Lendl},
  {Anderson}, {Collier Cameron}, {Delrez}, {Doyle}, {Gillon}, {Hellier},
  {Jehin}, {Maxted}, {S{\'e}gransan}, {Smalley}, {Queloz}, {Pollacco},
  {Southworth}, {Tregloan-Reed}, {Udry}, \& {West}}]{triaud_2017}
{Triaud}, A.~H.~M.~J., {Neveu-VanMalle}, M., {Lendl}, M., {et~al.} 2017, \mnras

\bibitem[{Weiss \& Schlattl(2007)}]{Weiss_2007}
Weiss, A. \& Schlattl, H. 2007, Astrophysics and Space Science, 316, 99

\bibitem[{Winn(2010)}]{Winn_2010}
Winn, J.~N. 2010, ArXiv e-prints, arXiv:1001.2010v5

\bibitem[{Winn {et~al.}(2008)Winn, Holman, Torres, McCullough, Johns-Krull,
  Latham, Shporer, Mazeh, Garcia-Melendo, Foote, Esquerdo, \&
  Everett}]{Winn_2008}
Winn, J.~N., Holman, M.~J., Torres, G., {et~al.} 2008, ApJ, 683, 1076

\end{thebibliography}

\appendix
\section{Radial-velocity data}
\begin{table*}[htbp]

\centering
\caption{CORALIE radial velocities of EBLM J0555-57A, and the probability that a point is not contaminated by the blend star. The radial velocity data are seperated into two sets, before and after t$_{\rm JDB-2,400 ,000}$ = 56770.48, to account for an upgrade of the instrument \citep{triaud_2017}. Separate systemic velocities are used for each set, determining $\gamma_1 = 19.537 \pm 0.015$ km s$^{-1}$, $\gamma_2 = 19.491 \pm 0.008$ km s$^{-1}$. All times are given in the BJD$_{\rm UTC}$ timestamp.}

\begin{tabular}{rrrrrrrl} \hline \hline
	
BJD - 2 400 000 & RV & 1$\sigma$ & FWHM & Bisector span & Depth of normalized CCF & Probability & Spectrum rejected \\
day & km s$^{-1}$ & km s$^{-1}$ & km s$^{-1}$ & km s$^{-1}$ & \% & & \\ \hline

56610.686754 & 25.078 & 0.012 & 14.179 & 0.310 & 17.149 & 0.000 & x \\

56615.797803 & 20.914 & 0.014 & 13.039 & 0.242 & 18.309 & 0.000 & x \\

56629.624924 & 13.125 & 0.018 & 14.624 & 0.101 & 16.583 & 0.000 & x \\

56640.778888 & 25.616 & 0.022 & 14.974 & 0.621 & 16.129 & 0.000 & x \\

56641.641065 & 25.229 & 0.020 & 14.362 & 0.424 & 16.560 & 0.000 & x \\

56644.700696 & 12.230 & 0.022 & 15.307 & 0.262 & 15.542 & 0.000 & x \\

56695.539666 & 26.056 & 0.021 & 14.697 & 0.215 & 16.161 & 0.000 & x \\

56697.572545 & 16.661 & 0.022 & 13.917 & -0.463 & 16.930 &  N/A &  x\tablefootmark{a}\\ 

56715.565737 & 15.302 & 0.014 & 13.869 & -0.056 & 17.264 & 0.560 &  \\

56716.588522 & 20.783 & 0.014 & 13.690 & 0.069 & 17.392 & 0.082 &  \\

56717.511413 & 24.949 & 0.023 & 13.837 & 0.071 & 16.637 & 0.873 &  \\

56718.547982 & 26.902 & 0.023 & 13.938 & 0.081 & 16.731 & 0.183 &  \\

56719.534391 & 24.542 & 0.023 & 13.828 & 0.215 & 16.913 & 0.962 &  \\

56723.625576 & 17.069 & 0.018 & 13.624 & -0.201 & 17.246 & 0.013 &  \\

56739.593653 & 19.399 & 0.015 & 13.797 & 0.012 & 16.991 & 0.726 &  \\

56740.590941 & 24.250 & 0.017 & 13.809 & 0.051 & 16.914 & 0.847 &  \\

56744.513089 & 13.487 & 0.020 & 13.866 & 0.073 & 16.544 & 0.610 &  \\

56746.538624 & 15.208 & 0.021 & 13.870 & -0.086 & 16.822 & 0.577 &  \\

56770.473218 & 19.968 & 0.021 & 9.1723 & 0.021 & 26.673 & N/A & x \tablefootmark{b} \\

56770.488252 & 18.650 & 0.020 & 13.668 & -0.045 & 17.040 & 0.059&  \\

57082.579230 & 25.648 & 0.032 & 14.406 & 0.309 & 17.435 & 0.000 & x \\

57085.612435 & 14.950 & 0.023 & 13.866 & -0.022 & 18.013 & 0.618 &  \\

57086.640474 & 11.384 & 0.022 & 14.154 & 0.338 & 17.643 & 0.000 & x \\

57117.510586 & 11.564 & 0.032 & 14.182 & 0.158 & 17.501 & 0.000 & x \\

57400.680862 & 26.035 & 0.030 & 14.424 & 0.260 & 17.408 & 0.000 & x \\

57417.703102 & 24.652 & 0.025 & 13.768 & 0.079 & 18.222 & 0.527 &  \\

57420.648080 & 11.791 & 0.021 & 13.786 & 0.113 & 18.130 & 0.648 &  \\

57422.573335 & 21.016 & 0.026 & 13.674 & 0.067 & 18.352 & 0.090 &  \\

57772.591031 & 25.051 & 0.035 & 13.850 & 0.086 & 18.070 & 0.797 &  \\

57773.731230 & 26.807 & 0.025 & 13.916 & 0.000 & 17.930 & 0.296 &  \\

57774.695330 & 23.836 & 0.033 & 13.777 & 0.216 & 18.015 & 0.626 &  \\ \hline

\end{tabular}%
\tablefoot{\\
	\tablefoottext{a}{Rejected before the FWHM outlier analysis due to anomalous span of the bisector inverse slope.}\\
	\tablefoottext{b}{Spectrum of EBLM J0555-57B}
}
\label{tab:addlabel}%

\end{table*}%

\end{document}